\documentclass[10pt,journal,epsfig,twoside]{IEEEtran}
\usepackage{amsmath}
\usepackage{graphicx}
\usepackage{amssymb}
\usepackage{amsfonts}
\usepackage{amsthm}
\usepackage{cite}
\usepackage{bm,comment}
\usepackage{algorithm}
\usepackage{algpseudocode}

\usepackage{subeqnarray}
\usepackage{subfigure}
\usepackage{multirow}
\usepackage{stfloats}
\usepackage{float}
\usepackage{color} 
\usepackage{cases}

\begin{document}
\title{Intelligent Reflecting Surface Assisted Terahertz Communications toward 6G}

\author{ Zhi Chen, Boyu Ning, Chong Han, Zhongbao Tian, and Shaoqian Li, \IEEEmembership{Fellow, IEEE}

%\thanks{Zhi Chen, Boyu Ning, Zhongbao Tian, and Shaoqian Li are with the University of Electronic Science and Technology of China. Chong Han is with the Shanghai Jiao Tong University, China. Boyu Ning is the corresponding author.}
}

\maketitle

\begin{abstract}
Terahertz (THz) communications have emerged as a promising candidate to support the heavy data traffic and exploding network capacity in the future 6G wireless networks. However, THz communications are facing many challenges for practical implementation, such as propagation loss, signal blockage, and hardware cost. In this article, an emerging paradigm of intelligent reflecting surface (IRS) assisted THz communications is analyzed, to address the above issues, by leveraging the joint active and passive beamforming to enhance the communication quality and reduce overheads. Aiming at practical implementation, an overview of the currently available approaches of realizing THz active/passive beam steering at transmitter and IRS is presented. Based on these approaches, a beam training strategy for establishing joint beamforming is then investigated in THz communications. Moreover, various emerging and appealing 6G scenarios that integrate IRS into THz communications are envisioned. Open challenges and future research directions for this new paradigm are finally highlighted. 

\end{abstract}

\section{Introduction}
The development of wireless communication systems is always in pursuit of higher data rates and system capacity while reducing latency and power consumption. Over the past few decades, the regulators in industry standardization and researchers from academia have been actively exploring approaches to improve communication quality. For the next generation system 6G, the wireless communication systems are expected to extend the communication spectrum to the Terahertz (THz) band ($0.1$-$10$ THz), which is expected to break the current bandwidth bottleneck compared to 5G systems under $100$ GHz\cite{FT}. Since 2008, a THz Interest Group (IGthz) has been established under the 802.15 umbrella, with the standardization of THz wireless communications by the IEEE. Hitherto the International Telecommunication Union (ITU) and World Radiocommunication Conference 2019 (WRC-19) have listed spectrum allocations up to $158$ GHz within the frequency range $252$-$450$ GHz\cite{KR}. 

\begin{figure}[t]
\centering
\includegraphics[width=3.5in]{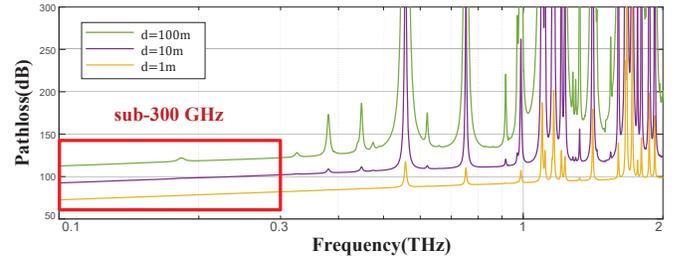}
\caption{Propagation loss in THz communication for different values of the transmission distance $d$.}\label{loss}
\end{figure}

There is no such thing as a free lunch. Although THz communications can provide a much more available spectrum band, it meets challenges from the wireless propagation loss, the signal coverage, and the antenna/radio-frequency fabrication\cite{IF}. Specifically, i) Friis' transmission formula indicates that doubling the frequency quadruples the propagation loss. In addition, the molecular absorption starts to significantly impact the propagation loss in THz communication, especially at a frequency around $60$ GHz (oxygen) and above 300 GHz (water vapor). As shown in Fig. \ref{loss}, the THz propagation spectral lines are separated by the molecular absorption peaks into multiple transmission windows, which is a notable distinction compared to the mmWave counterpart. ii) The diffraction and scattering ability of the THz wave is very poor. Thus, THz signals have the property of high directivity are easily blocked by obstacles in the wireless environment. Thus, the line-of-sight (LoS) transmission is mainly considered in THz communication, which leads to limited signal coverage. Regarding this point, the coverage enhancement should be given primary consideration. iii) In 5G systems, the multiple-input and multiple-output (MIMO) technologies have been well-developed to increase the beam gain for compensating propagation loss. When utilizing the THz band, the antenna array can be packed in a small footprint with a large number of elements, which is attributed to its tiny wavelength and high transmission gain requirement. However, manufacturing THz components, with PIN diodes, varactors, triodes, and so on, in small size is very difficult and of high cost. As a result, the widespread use of THz communication might be hindered if the above issues cannot be well addressed.
\begin{figure*}[t]
\centering
\includegraphics[width=7in]{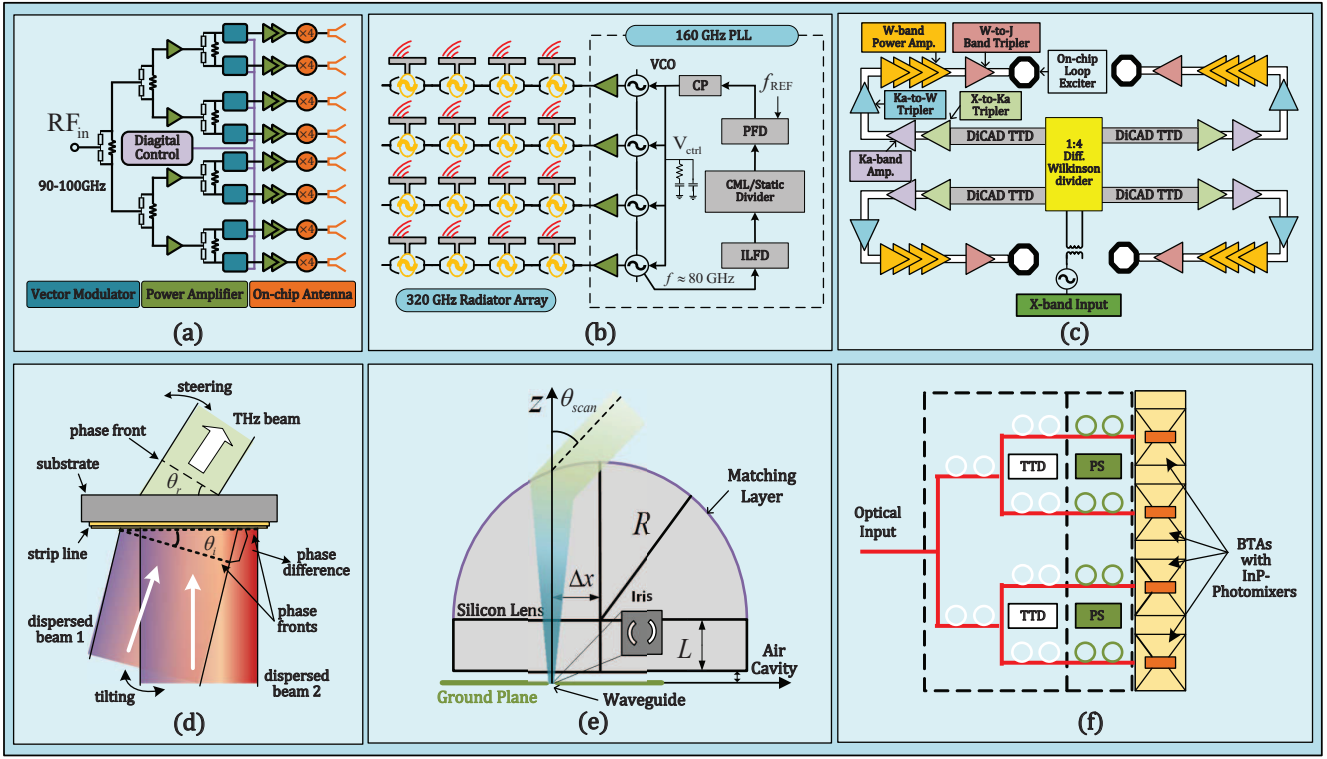}
\caption{Approaches of THz active beam steering at transmitter: a) modulating the phase in the lower frequency and then convert to the THz frequency\cite{yyang}; b) directly modulating the phase in the THz band\cite{rhan}; c) using the digital controlled TTD antenna array\cite{nub}; d) changing the phase difference of two laser beams\cite{kmc}; e) refraction based on the optical lens antennas\cite{lens}; f) using the optical TTD-based antenna array\cite{plu}.}\label{transm}
\end{figure*}

Recently, a new concept called intelligent surfaces has emerged, which provides an interface between the physical electromagnetic world and the digital world. The intelligent surface, also referred to as metasurface, is composed of artificial unit cells possessing different electromagnetic (EM) responses, providing new possibilities to realize the real-time manipulation of EM waves\cite{MD}. In general, there are two types of intelligent surfaces, i.e., intelligent transmission surface and intelligent reflecting surface (IRS). For the former, the incident EM wave is converted into the transmitted signals via going through the intelligent surface with its amplitude and phase adjusted by the external control signals. In contrast, the latter is more appealing wherein the intelligent surface is mounted on the environmental objects and the incident EM wave will reflect via it with adjustable responses, which enables the wireless environment to be controlled. As the conventional THz communication technologies meet challenges in the propagation loss, the LoS blockage, and the hardware cost, integrating the IRS into THz communications provides a promising low-cost solution to alleviate the short-range bottleneck and establish reliable wireless connectivity. Specifically, with the assist of IRS, joint active and passive beamforming can be realized in the THz communication to enhance spectrum and energy efficiency, and provide virtual LoS paths to reduce the probability of blockage\cite{huangchi}. 
In light of this, IRS-assisted THz communication is considered a promising paradigm in the future 6G systems. In this article, we first overview the joint beamforming technologies from the perspectives of both the hardware design and the software-defined manner in THz-IRS communications. Next, we present some emerging scenarios of THz-IRS communications and illuminate their advantages in 6G applications. Then, we discuss the main challenges and the promising future research directions. Finally, we conclude the article.

\section{Joint Beamforming Technologies}
Beamforming is the focus of interest in THz-IRS communications wherein the core problem lies in how to steering signal energy to the desired targets. Traditional MIMO arrays and IRS devices are hard to scale at the THz frequencies. Thus, an urgent concern is how to design the hardware with THz beam steering capability. In this section, the state-of-the-art THz beam steering approaches at transceiver and IRS are reviewed, respectively. Then, we further introduce an IRS-assisted beam training strategy, which is used to establish joint beamforming in THz-IRS communications.

\subsection{Active Beam Steering at THz Transmitter}
The THz transmitter generates directional electromagnetic radiation through its own feed, which is usually called \emph{active beamforming}. Conventional directional antennas, such as horn and Cassegrain reflector antennas can be commercially found at frequencies of up to $1$ THz. The simplest way to achieve the beam steering via these antennas is to mechanically rotate them, which has been employed in military radar systems. However, it is not viable for modern communications applications owing to the latency requirements and power consumption. In the following, we list some prospective THz transmitters that can realize tunable beam steering.

\subsubsection{Electronic Approaches}
As the efficacy of electronic components is constrained in the THz band, a feasible solution is to modulate the phase in the lower frequency and then convert to the THz region. As such, the authors in \cite{yyang} propose a $370$-$410$ GHz $8\times 1$ uniform linear array (ULA), the architecture of which is shown in Fig. \ref{transm} (a). It is observed that the $90$-$105$ GHz signal is divided into $8$ channels by Wilkinson splitters, with the phase modulated by the vector modulator. After that, the signal is fed into the W-band amplifiers. At last, the frequency of the signal is converted to $360$-$420$ GHz by the quadrupler. The up-conversion method avoids the difficulty of phase shifts in the THz band, which however cannot allow complex high-order modulation signals. 

Apart from this approach, Adler's equations provide a straightforward way to directly modulate the phase in the THz band via injecting an external signal into the oscillator. Based on it, the authors in \cite{kse} present a scalable $4\times 4$ uniform planar array (UPA) in CMOS at $280$ GHz, wherein the power generation and radiator core are based on distributed active radiation (DAR).  Another example is given in \cite{rhan}, where a $4\times 4$ UPA in SiGe at $320$ GHz was proposed based on coupled harmonic oscillators. As shown in Fig. \ref{transm} (b), this array is made up of $16$ elements and a fully-integrated $160$ GHz phase-locked loop (PLL), consisting of charge pump (CP), phase/frequency detector (PFD), current-mode logic (CML), and injection-locking frequency divider (ILFD). Directly modulating the phase at THz frequencies provides more flexibility for beam control, which unfortunately comes at the cost of high hardware complexity.

An alternative attractive way is to replace the phase shifters by the digital controlled true time delay (TTD) elements. The authors in \cite{nub} propose a $280$ GHz $2 \times 2$ chip-scale dielectric resonator antenna array. As shown in Fig. \ref{transm} (c), this array incorporates a balun for an X-band input signal, a $1:4$ Wilkinson divider, and four x$27$ active multipliers chains to drive the elements, where the delay of each chain is controlled by a digitally controlled artificial dielectric transmission lines (DiCAD)-based TTD. As the TTD can compensate for signal delay at all frequencies, it can eliminate the beam squint problem in wideband communication.

% \textcolor{red}{Despite that the electronic approaches are limited by radiated power and processing bandwidth, they provide compact integration and high beamforming flexibility to miniaturized THz communications.}

\subsubsection{Optical Approaches}
If two laser beams are focused onto a photoconductive antenna, by changing the phase difference of the two dispersed beams, the THz beam can be generated with adjustable direction\cite{kmc}. Fig. \ref{transm} (d) illustrates the principle of steering THz beam via pumping beam tilt. Specifically, two optical beams impend to the surface of the stripline from different directions. $\theta_i$ is the angle variation between the two pump beams, which results in the graded distribution of the phase difference. The phase difference $\theta_r$ of the generated THz wave is proportional to that of the two incident beams, which leads to the adjustable THz beam steering. This method avoids the huge phased array structure and expensive phase shifters.

Based on the optical lens antennas, refraction is also a classical way to change the direction of the THz beams. The authors in \cite{lens} demonstrate the beam steering capability of a silicon lens antenna fed by a THz wave feed at $550$ GHz. As shown in Fig. \ref{transm} (e), the antenna consists of a leaky waveguide feed that illuminates only the upper part of a hemispherical lens. The feed is composed of a square waveguide, an iris on a ground plane, and an air gap between the iris and the silicon lens. The lens can be moved driven by a piezoelectric motor. The movement of the lens relative to the waveguide nominal position leads to the change of the beam direction. The lens made of non-dispersive material supports extremely high bandwidth, which is advantageous for THz wideband communication. However, it cannot support high-speed and high-precision beam steering, due to the limitation of the travel speed and movement accuracy of the piezoelectric motor.

Recently, the optical TTD phase shifters (PSs) are also employed to offer stable time delay for wideband communications\cite{plu}. Fig. \ref{transm} (f) presents the schematic view of an optical TTD-based chip, wherein the input optical signals will be converted to $300$ GHz frequency region by the InP photomixers and finally radiated by the $1 \times 4$ bow-tie antenna. The quality of the input laser source in the TTD-based array would noticeably affect the practicability of the generated THz beam.

\subsection{Passive Beam Steering at THz IRS}
IRS consists of arrays of metamaterial, which can be used to precisely control incident electromagnetic fields, so as to achieve \emph{passive beamforming}. The word ``passive'' indicates that there is no electromagnetic wave generated from IRS. However, there need some low-cost active components to control the elements for adjusting the reflected electromagnetic fields. The semiconductor lumped components widely used in microwave IRS are not appealing in the THz band due to their fabrication challenges and unaffordable cost. Fortunately, microelectromechanical systems (MEMS) and some tunable and reconfigurable materials, such as graphene, liquid crystal (LC), transparent conducting oxide, vanadium dioxide (VO2), 2D electron gas, halide perovskite, quantum dot, and superconductor, can be utilized to realize IRS at THz frequencies benefiting from their freedom of regulating amplitude response, phase response, and polarization. The excitation method is no longer limited to electric control, and the thermal environment, as well as the light pump, also become effective incentives. In the following, we introduce two main electrically tunable materials, i.e., graphene and LC, which interest communication engineers most.
\begin{figure}
\centering
\includegraphics[width=3.1in]{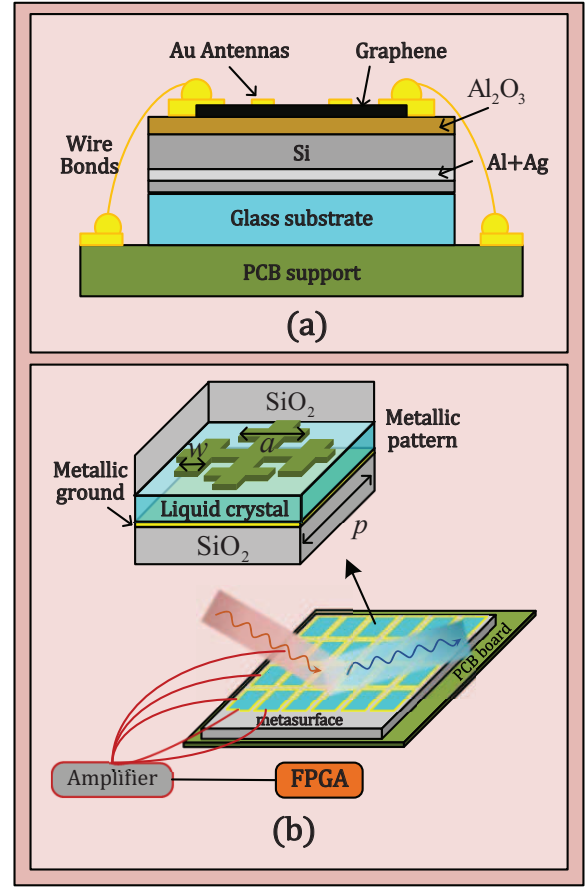}
\caption{Approaches of THz passive beam steering at IRS: a) graphene-based THz IRS\cite{mta} ; b) LC-based THz IRS \cite{jwu}.}\label{irsf}
\end{figure}

\subsubsection{Graphene-based THz IRS}
Graphene is ideally suited to modulate THz waves owing to its rich physics and gate tunable properties. Several experiments have been carried out and some promising results have been validated. For instance,  the authors in \cite{mta} experimentally demonstrate a $0.98$ THz one-bit programmable graphene reflectarray metasurface that can achieve THz beam steering. As shown in Fig. \ref{irsf} (a), the reflectarray is mounted on a printed circuit board (PCB) and wire-bonded, where the substrate is comprising a reflective conductive ground plane and a dielectric spacer. Each element can exhibit opposite reflection phases via two different biased voltage, i.e., ${\mathrm{V_{on}}}$ and ${\mathrm{V_{off}}}$. 
The reflectarray contains $400 \times 80$ cells, wherein the $400$ cells in each column are controlled by the same bias voltage. This simplified control network makes the IRS behave as a $80$-element linear array with one-bit coding. The steering range of the deflection angle can reach almost $25$ degrees by setting different phase combinations on these elements. When the scale of 2D graphene IRS grows to a large scale, how to design a huge and complicated driving network to quickly control a tiny IRS remains an inevitable challenge. There is an urgent need for more experimental verification of graphene IRS in the THz band.

\subsubsection{LC-based THz IRS}
By leveraging the birefringence effect of LC, the phase retardation can be altered by switching the orientation of the LC molecules dynamically. Thus, the LC-based IRS enables the
dynamic control of THz beam deflection. The authors in \cite{jwu} demonstrate a $0.67$ THz LC reflectarray with one-bit programming capability. As shown in Fig. \ref{irsf} (b), it consists of a $24$-element linear array, and each element is composed of $50$ rows and $2$ columns of unit cells with metal-insulator-metal resonator structure. When the electric field is applied between the two metallic layers, the LC molecules reorient, and the refractive index changes correspondingly. As such, similar to the manipulations in \cite{mta}, by using two different biased voltage to control the reflection phases on each element, beam steering can be realized. Based on this reflectarray, it has been proved that a maximum deflection angle of $32$ degrees can be achieved.

Compared to the graphene reflectarrays, most LC reflectarrays exhibits less sensitivity of beam control in practical experiments.  This is due to the fact that the change in the phase retardation of LC-based IRS relies on an electric field to reorient the molecules in the liquid, and the reorientation speed of the molecules limits the switching response time. By contrast,
the graphene reflectarray supports adjustment of the reflection phase with ultra-fast switching delay, i.e., on the order of the picosecond.

% \textcolor{red}{Since the composite conductivity of the graphene patch controlled by the electric field bias depends on the rearrangement of the carriers near the speed of light.
% In summary, the random pointing THz beam access puts forward higher requirements for electronically controlled precise beam steering. Both active/passive beamforming methods should seriously consider this issue.}

\subsection{Joint Beamforming via Beam Training}
\begin{figure}[t]
\centering
\includegraphics[width=3.5in]{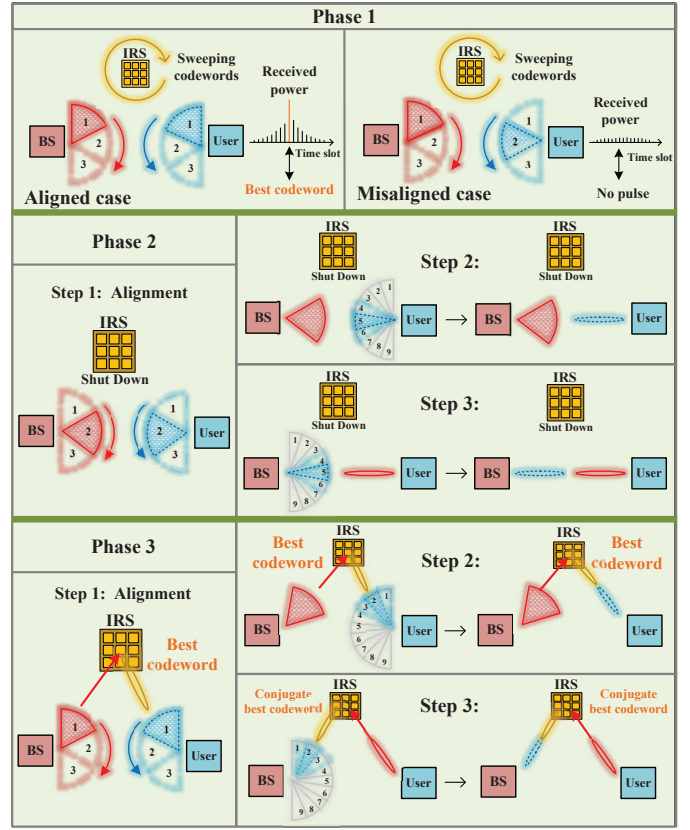}
\caption{The procedure of beam training in THz-IRS systems\cite{nby}.}\label{tran}
\end{figure}

Benefiting from the steering ability at the THz transceivers and IRS, a software-defined manner can be used to connect the beams and establish reliable communication, which is known as a \emph{beam training} strategy. As IRS can not generate and decode beams, beam training in IRS-assisted systems is much more challenging than that in conventional scenarios. To realize a joint beamforming scheme for IRS-assisted systems, three phases are needed to achieve different groups of beam alignments\cite{nby}. Considering a base station (BS), a user, and an IRS, the basic principle of the beam training is shown Fig. \ref{tran}, and also outlined below.
\begin{figure*}[t]
\centering
\includegraphics[width=7in]{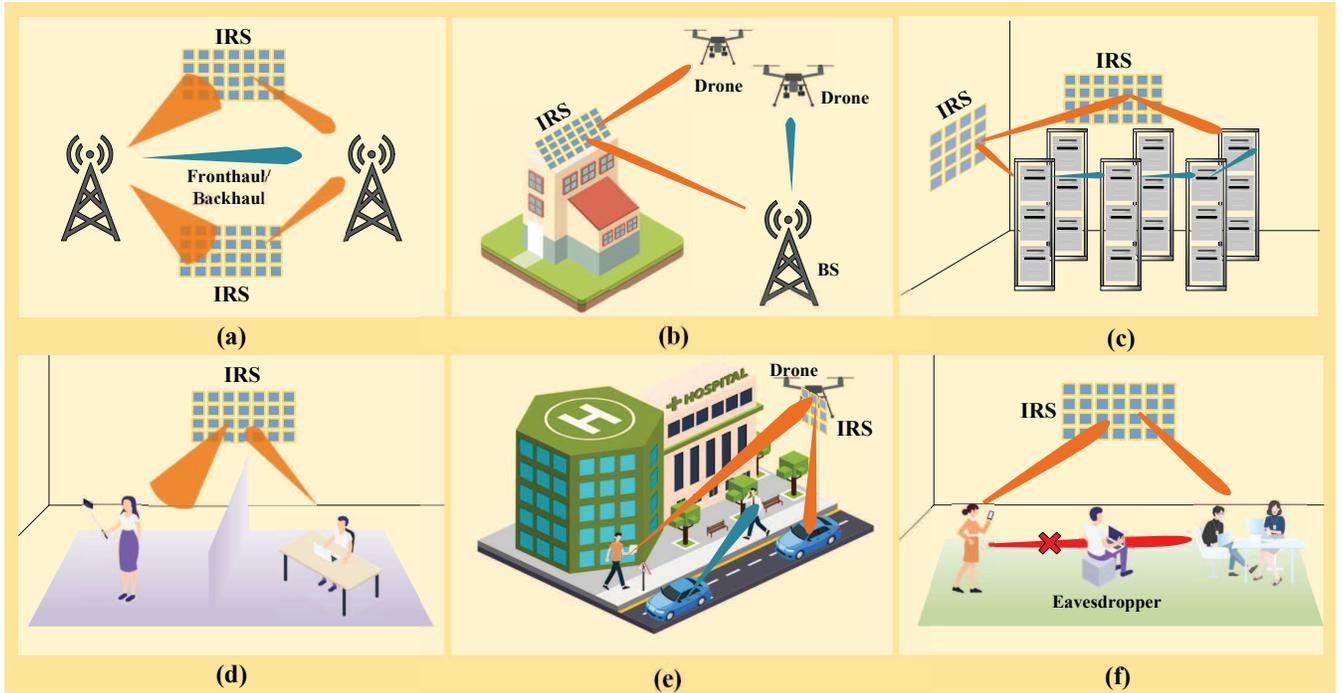}
\caption{Promising applications of THz-IRS communications in six scenarios: a) high speed fronthaul/backhaul; b) cellular connected drones; c) wireless data center; d) enhanced indoor coverage; e) vehicular communication; f) physical layer security.}\label{scenar}
\end{figure*}
\subsubsection{Phase 1}
We aim to obtain the best codeword (i.e., the optimal reflecting mode) for IRS. First, we test some wide-beam pairs in successive intervals with BS using the transmitting mode and user using the receiving mode. In each interval, the IRS successively searches predefined codewords in different time slots. For the IRS, there is only one wide-beam pair that covers both the BS-IRS link and the IRS-user link. During this interval (aligned case), the user will detect an energy pulse in the time slot when IRS uses the best codeword. Thus, the user can utilize the pulse slot to identify the best codeword for IRS.
\subsubsection{Phase 2}
We turn off the IRS and obtain the BS-user narrow-beam pair via the following three steps. In step 1, some wide-beam pairs are tested for coarse alignment. The user compares the received energy in different intervals and determines the wide-beam pair with the maximum power. The aligned wide-beam pair will be labeled on both sides. In step 2, the BS transmits the labeled wide beam and the user uses a fast hierarchical search to find its optimal narrow beam. In step 3, the user transmits this narrow beam and BS uses a fast hierarchical search to obtain the best narrow beam at BS.
\subsubsection{Phase 3}
We aim to obtain the BS-IRS-user narrow-beam pair through three steps similar to Phase 2. We turn on IRS with the obtained best codeword. There exist two propagation paths from the BS to the user, i.e., the BS-IRS-user path and the BS-user path. Note that the BS-user path has been determined in Phase 2. Hence, we can find the narrow-beam pair for the reflecting path by a fast hierarchical search as long as removing the received signals along the BS-user path. By this means, we can obtain the best narrow-beam pair for the BS-IRS-user path. 

As the IRS provides a virtual LoS link, i.e., through the reflecting link, the probability of LoS blockage is significantly reduced. Furthermore, one can mount multiple IRSs on different locations to further increase the signal coverage. As long as there is one unblocked link, the connection is reliable\cite{nby}. We can reduce the beamwidth of the wide beams used for alignment if additional beam gain is demanded in the initial link connection. This because the narrower the beamwidth, the higher the beam gain. However, this comes at the increasing cost of complexity for alignment.

\section{EMERGING SCENARIOS}\label{usecase}
In this section, we present some appealing scenarios of THz-IRS communications in 6G applications. Benefiting from the unique function of IRS, the THz communication scenarios in 6G will be more diversified.

\subsection{High Speed Fronthaul/Backhaul}
The 6G wireless networks are envisioned to use the dense deployments of cellular cells to meet the unprecedented data rate requirements of hotspots. As envisioned by the IEEE 802.15.3d standard, through beamforming of large scale antennas to provide high gain transmission, THz ultra-high speed wireless communication can be used to flexibly deploy fronthaul (between baseband units and remote radio units) and backhaul (between cells and the core network) links, thereby reducing the complexity and cost issues caused by wired optical fiber connections. More than that, THz communication can also use the widely deployed IRS for joint beamforming to provide high throughput fronthaul/backhaul transmission, as shown in Fig. \ref{scenar} (a). The IRS-assisted THz wireless fronthaul/backhaul link not only provides extra aperture gain through passive beamforming at the IRS but also reduces outages by establishing multiple propagation paths.

\subsection{Cellular Connected Drones}
Drones can be regarded as virtual base stations or mobile relays that support flexible deployment to provide adjustable large scale coverage for 6G networks. To fully utilize the potential of drones, 6G is expected to use the THz band to support heavy traffic between drones and BS as well as users. However, the openness of the air space makes THz communications that rely on high power beam to combat fading easily to cause mutual interference, which in turn hurts the transmission performance. As an easy way to adjust the reflection angle of electromagnetic waves, IRS can intelligently adjust the direction of the beam from the BS to provide a variety of drone connections, as shown in Fig. \ref{scenar} (b). In urban areas, IRS is generally recommended to be deployed on the exterior wall or roof of a building. Introducing IRS into the airspace BS beamforming scene can improve the interconnection capability of air ground communication while avoiding multi drones interference.

\subsection{Wireless Data Center}
As the mobile communications demand for cloud service applications increases steadily, data centers will play a more important role in 6G. With its higher reconfigurability and dynamic operation, THz enabled wireless data centers have the potential to solve the defects of power consumption, maintenance costs, and space occupied by large cables in wired data centers. However, the point-to-point LoS link established between densely arranged data servers will inevitably be blocked by the server itself or cause link interference. Therefore, as shown in Fig. \ref{scenar} (c), the introduction of IRS into the THz wireless data center to establish IRS-assisted THz links will greatly expand the freedom of server interconnection path planning. On the one hand, IRS can provide a variety of link route options for quasi-static data centers. On the other hand, IRS can assist in the establishment of multiple THz backup connections to improve the reliability of data center transmission.

\subsection{Enhanced Indoor Coverage}
Due to the high frequency and short wavelength, the high transmission loss of THz communication results in a short coverage distance, while the limited diffraction capability causes THz transmission to rely on the LoS path. Indoor THz LoS link can be easily blocked by the walls or human bodies, leading to high speed communication interruption. For the sake of human health and energy-efficient communication, indoor THz wireless link cannot fill coverage holes by simply increasing the transmission power or setting up more access points. As such, it is an arduous task to expand the ubiquitous high-speed indoor coverage in THz communication.  IRS has become an innovative and cost-effective THz indoor coverage solution due to the following main reasons: i) IRS can provide a virtual wireless LoS link by controlling the reflection angle. ii) IRS does not need complex hardware circuits and has a small thickness with lightweight. These physical characteristics enable the IRS to be easily installed in wireless transmission environments, including walls, ceilings, and furniture. An example is sketched in Fig. \ref{scenar} (d), where the IRS assists the THz access point to establish a virtual LoS path behind the obstacle to enhance indoor coverage.

\subsection{Vehicular Communications}
The intelligent transportation system expects the vehicle communication network to provide high data rate, low latency, and reliable communication. For the era of wireless interconnected smart cars, THz communication is a potential support technology for terabit vehicular communications in the future. However, the changeable crowded traffic and dense movement of people will damage the connection stability and alignment speed of the THz beam. To this end, the mobile IRS carried by the drones can follow the traffic flow to assist the THz beam training and tracking process in crowded traffic areas. Fig. \ref{scenar} (e) shows a schematic diagram of the ultra-high speed THz link used for vehicle communication, in which the IRS carried by the drone can be adjusted to different heights and positions as required. Vehicles can choose cooperative IRS according to obstacle conditions at different locations to ensure high-speed, real-time, and stable THz connection.

\subsection{Physical-Layer Security}
As wireless network security relying solely on high-level encryption protocols remains limitations, it is significant to consider physical layer security in 6G networks. Using massive antenna elements to create highly directional THz beams brings many benefits to the secure physical layer transmission. However, eavesdroppers located in the sharp sector of the beam coverage can still endanger information security. By jointly active and passive beamforming, the IRS-assisted THz communication systems can concentrate the beam energy to the legitimate user while suppressing the eavesdropper's received power. An illustrative example is sketched in Fig. \ref{scenar} (f), in which IRS not only helps the THz beam to bypass the eavesdropper in a reflection path but also intentionally deteriorates the signals power in the direction of the eavesdropper.

\section{Challenges and Future Research Directions}
Although THz-IRS communications are expected to bring a major leap for 6G systems, there are still many essential problems in implementing efficient and pragmatic applications. In this section, we outline some main challenges and future research directions.

\subsection{Transceivers and IRS Hardware Design}
Most of the existing THz antenna and IRS only consider single carrier modulation and the array scale is very small, i.e., less than $20$ elements. While realizing the wideband communication via ultra-massive MIMO, the following challenges needed to be considered. First, the design of THz ultra-massive MIMO antenna arrays faces the test of component feed wiring and large array heat dissipation. Second, the nonlinear effect of THz radio frequency devices makes it difficult to efficiently amplify THz signal, making the entire THz systems face the predicament of high energy consumption and low transmission power. Third, due to the short wavelength of the THz wave, the subwavelength size of IRS increases the number of integrated reflective elements per unit area, e.g., $1024$ elements in $1$ mm$^2$ at $1$ THz. Thus, the combination of a large number of reflective elements and semiconductor devices becomes difficult. Forth, tunable-material based IRS, such as VO2, LC, and graphene, generally have different responses at different frequencies, which makes the design of passive beamforming more complicated for wideband communication. Given the dilemma encountered by THz transmitter, the reflectarray antennas activated by IRS might replace traditional transmitters to realize waveform control, modulation, beamforming, etc.

\subsection{Theoretical Propagation Modeling}
As the THz-IRS systems adopt brand new materials and hardware designs, it is necessary to develop theoretical propagation model considering both the reflection attenuation and response mode of the THz wave on the IRS elements. From the perspective of signal characterization and electromagnetic response, the following points need to be concerned.
First, the impact of the THz ultra-massive MIMO antenna on the near-field range cannot be ignored. 
Second, non-ideal IRS hardware response, e.g., the effect of temperature on reflection characteristics and phase response with errors, should be considered at the theoretical level.
Third, the IRS reflection propagation model should be separately established for different THz sub-bands according to their unique diffraction/scattering ability and molecular absorption.
Forth, there is still a lack of intelligent programmable wireless environments in the THz band to test the effectiveness of channel modeling through practical measurements.

\subsection{Beam Management}
The ultra-massive MIMO THz systems utilize extremely narrow beams to combat the severe propagation loss, which requires the beam to point to the receiver accurately. The emerging scenarios of IRS-assisted THz communications, e.g. cellular-connected drones and vehicular communications, further require beams with the capability of 3D steering. Moreover, the THz-IRS systems need to support high-speed mobile transceivers in the air and ground, which brings challenges in beam management.
Therefore, the challenges related to beam management are enlightened as follows.
First, the 3D codebook designs are needed for the practical THz transceivers and IRS, e.g., with hybrid beamforming architecture or low-bit quantized phase shifters. Second, the beam training and beam tracking technologies require the co-design of the transmitter and the IRS to support the high-speed mobile transceivers.
Third, the beam scheduling and the interference suppression need to be considered at both the transmitter and IRS in multi-user systems.

\subsection{Channel Estimation and Transmission Optimization}
Obtaining accurate channel state information (CSI) is an essential condition of optimal beamforming. With the CSI, flexible beamforming can be realized by optimizing the weights of amplitude and phase on each antenna element, rather than simply steering a single beam. However, some challenges are needed to be considered.
First, the conventional pilots generally do not have adequate beam gain. Thus, it may be detected with a lower SNR, or even cannot be effectively detected by the receiver, owing to the severe propagation loss in THz communication. The pilots with low reception quality make the estimation of CSI inaccurate, which puts forward higher requirements on the robustness of the joint beamforming algorithm.
Second, the ultra-massive MIMO THz systems contain thousands of elements, which raises the estimation overhead dramatically. Thus, fast channel estimation algorithms are required to meet the needs of low-latency applications in 6G, which is a trade-off between the time cost and estimation accuracy.
Third, as IRS cannot process or transmit THz signals by itself, it is hard to estimate the transmitter-IRS channel and the IRS-receiver channel directly. The transceivers and IRS should achieve the channel estimation for the cascaded reflecting link in a cooperative manner.

\section{Conclusion}
In this article, the transformative features of IRS are considered to be incorporated into THz wireless communications, for realization of cost-effective and energy-efficient 6G networks. Based on the existing electronic and photonic approaches, THz beams can be excited with tunable capability. Furthermore, new metamaterials, such as LC and graphene, are promising technologies to support the control of THz beams.  Benefiting from these technologies, various emerging scenarios of IRS-assisted THz communications in 6G applications are envisioned to be realized in the near future. However, the implementation of many technologies is still in the early stage with open research issues. Unleashing the full potential of this paradigm requires efforts of researchers to provide enabling solutions for future IRS-assisted THz communication systems.

%}

%\section{Acknowledgments}
%The work is supported by the National Key Research and Development Project of China under Grant 2018YFB1801500. 


%\begin{IEEEbiography}{Michael Shell}Biography text here.
%\end{IEEEbiography}


%\newpage



%\vfill


\end{document}